# Reducing parasitic resonances in particle accelerators components by broadband Higher–Order–Mode couplers

P. Arpaia[a], O.E. Berrig[b], L. De Vito[c], A. Gilardi[a,b]

[a]*Instrumentation and Measurement for Particle Accelerators Laboratory, University of Naples Federico II, Naples, Italy*
[b]*Beam Department, CERN, Geneva, Switzerland*
[c]*Engineering Department, University of Sannio, Benevento, Italy*

**Abstract**

In particle accelerator components, parasitic resonances must be reduced because they heat up the equipment and cause beam instabilities. In this paper, a method for designing and characterizing Higher Order Mode (HOM) couplers for reducing such resonances in a broad bandwidth is proposed. A case study is considered for a specific component, called QuattroTank, showing geometrical discontinuities and thus causing significant electro-magnetic resonances. Results of numerical simulation and experimental emulation prove the capability of the proposed method to reduce the peaks and the *Q*–factor of the resonances.

## 1. Introduction

In particle accelerators, when a bunch travels at a speed close to that of the light, spacial discontinuity and material imperfections produce *wake fields* [1]. These wake fields interact with the accelerator equipment, and, if a component has a spacial discontinuity, i.e. a non-uniform cross-section, an Electro-Magnetic (EM) resonance is created [2]. Such resonances introduce twofold main problems [3]: (i) they can absorb a huge amount of energy, and (ii) can generate interference, and thus beam instability. Heating and instabilities in high–energy accelerators have been studied since the late 1950s [4]. In fact, they are among the main factors that determine the ultimate performance of the accelerator.

However, beam instability is a much more serious issue than heating, because often causes an irreversible loss of beam. On the other hand, the loss of energy by heating can usually be compensated by cooling the equipment. Avoiding beam instabilities is such an important task that the European Organization for Nuclear Research (CERN) has a dedicated personnel section at this aim.

The electromagnetic interaction of a component with the beam is modeled by its impedance analogously as the Ohm's law [5]. In this framework, an Electro-Magnetic resonance is modeled as a beam impedance at a single frequency. The current is modeled by the circulating charged particles and the voltage drop represents the force acting on the particles. Therefore, the beam impedance creates forces acting on the beam like dissipative frictions that takes energy out of the beam. There are two main types of component impedances: arising from the geometry and by material properties. Ideally, impedance does not arise if the vacuum chamber has zero surface resistance and a uniform cross-section over the whole accelerator.

Several methods have been proposed to reduce the beam impedance of equipment in particle accelerators. All of them have issues, therefore a single method cannot be used in all cases. Most common are [6]: transition pieces, coating, serigraphy, and absorbing materials (see Section 2). Generally, only narrow peaks in the beam impedance spectrum are a concern for beam stability and beam induced heating.

The defining properties of a resonant impedance are the resonant frequency ($f_r$), the geometry factor ($\frac{R_{shunt}}{Q}$) and the quality factor ($Q$). For a certain type of devices, it is unavoidable to have a cavity present in the structure. In this case, the beam impedance is to be reduced differently. These properties depends on several parameters. Both $f_r$ and $\frac{R_{shunt}}{Q}$ are strongly determined by the geometry of the structure, and thus cannot be changed without a significant modification of the device, which may hinder the intended operation. Thus, one approach is to change the $Q$ of the resonance.

A well–known method of changing the $Q$ of a resonant cavity is to add a dispersive component, usually the ferrite because it absorbs the magnetic field, to the cavity volume [7] (see Section 2). If an existing device has too high beam impedance, limitations of both time and budget may require the use of retroactive solutions to reduce the beam impedances.

Higher Order Mode (HOM) couplers are devices that remove or dissipate unwanted RF energy from a cavity. Nowadays, these couplers are used mainly inside actual accelerating cavities, where other modes (unwanted excitation by the beam) than the main accelerating mode [8] have to be damped. In [9], HOM couplers were proved to be capable of damping unwanted resonances. However, a formal statement of the method, as well as the complete design and the characterization of such devices, were still missing.

In this paper, a formal method to reduce parasitic resonances in particle accelerator components using a new type of HOM couplers is proposed. This method is specially good as a retroactive solution. In Section 2, state-of-the-art methods of damping unwanted resonance in particle accelerators are reviewed. In Section 3, a method for designing and characterizing HOM couplers specific for reducing parasitic resonances is described. In Section 4, a case study on the component *Quattrotank* is presented. Then, in Section 5, the simulation analysis aimed at verifying the feasibility of the proposed device to damp the resonances is reported. Finally, in Section 6, the actual design of the device and its experimental assessment are described.

**2. State of the art**

As already mentioned, the most common methods for damping unwanted resonances in particle accelerators are [6]: *transition pieces*, *coating*, *serigraphy*, and *absorbing materials*. *Transition pieces* reduce geometric impedance by smoothing out the discontinuities along the beam path. They could be made by one or several pieces of conducting material to screen any spacial discontinuity. They can be rigid or movable. Most commonly, they are used for devices that require some mechanical degree of freedom (i.e. longitudinal or transverse movement), such as bellows. An example of RF–fingers in the Large Hadron

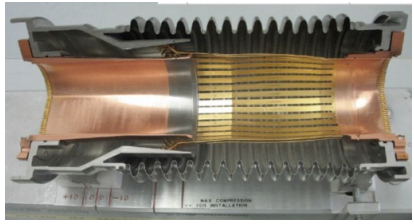

Figure 1: Example of RF fingers placed between magnet in the LHC.

Collider (LHC) magnets at CERN is shown in Fig. 1. These RF-fingers compensate for the longitudinal movement when the magnets are cooled down and become shorter. Transition pieces give an impedance reduction that is effective for a number of reasons. As an example, they provide a short, good conducting path for the image currents that otherwise would have flown into the cavity, and, in the case of bellows, the image current does not have to follow the irregular path of the bellow. The main drawback of this method is the sensitivity to mechanical failures. Mechanical stresses can easily break the transitions pieces; for example, in LHC in 2011, in 8 out of 10 of the double bellow modules, the springs holding the RF fingers were broken [10]. Another well-known method to reduce beam impedance is *coating* the structure by a material with a higher conductivity [11]. Normally, surface coating can be used on all types of equipment; however, sometimes operational constraints impose the material of the device. Typical examples of equipment that can not be coated are: kicker magnets and beam instrumentation. A further important issue is the dependency of the coating thickness on the frequency, which in turn is related to the bunch shape. As en example, the coating can not be used for coupled bunch instabilities, as this happens at low frequencies (around 8 kHz), while it is really efficient in reducing the resistive wall impedance. This phenomenon arises from the skin depth effect [12].

A not negligible issue for the methods presented so far is that both of them need costly physical modifications of the component.
Kicker magnets are a particular case. In kicker magnets, in order to reduce the eddy currents (caused by the fast ramping rate of the magnetic field) the vacuum chamber should be really thin. A drawback is that low–frequency image currents will penetrate the chamber walls and flow on the surface of the magnetic poles of the kicker magnet.

In this case, *serigraphy* is used to guide the image currents. It consists of printing a set of interleaved fingers on the magnet poles (see Fig. 2), made of a very good conductor (i.e. silver), to form a good conductive path for the beam image currents [13]. The interleaved fingers are coupled capacitively to each other. Main advantages of serigraphy are the minimal disruption to the device geome-try and its surfaces, and its low cost. The drawback is that often the equipment to be designed with serigraphy, because the small capacitive coupling, does not allow the flow of low frequency components of the current.

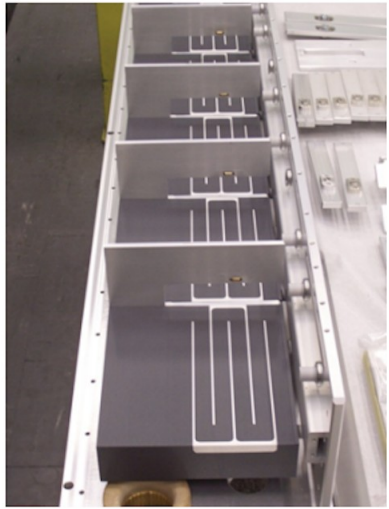

Figure 2: An example of serigraphy in the SPS Extraction Kicker Magnets (SPS-MKE).

Mainly ferrites are used as *absorbing materials* to damp cavity modes, which can not be removed by redesign, and hence are a subject of intensive study. However, since the ferrites absorb the energy of the resonance, they heat up and this is a big issue because they may heat beyond their Curie temperature, and then they will stop acting as dispersive material. In addition, they will start to out-gas [14]. These two problems, in certain cases, dramatically prevent the use of ferrites as damping material.

Classically, in the design of an actual accelerating cavity, it is already known that the beam excites unwanted modes, therefore HOM couplers are also designed and installed during the design phase [15]. These HOM couplers are quite big devices and the reason is that they needs to absorb the high energy of the unwanted modes that are trapped in the real accelerating structure [16]. These modes always have high Q-factors (therefore high energy) [17], because they are inside a cavity which is designed to have a high Q-factor.

In some cases, such modes are used to extract information useful for beam diagnostics, e.g. to determine the center of the structure [18] or the position of the beam [19].

Any HOM coupler can be composed by a loop or an antenna that interact with the magnetic or the electrical field (it depends on the configuration used), which characterizes the unwanted mode, the two setup are shown in Fig. 4. The power caught by the HOM coupler will be dissipated on a resistor that is outside the equipment.

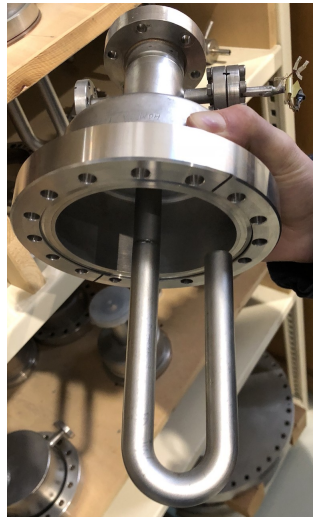

Figure 3: HOM coupler actually used in an accelerating cavity

The HOM coupler in the probe configuration interacts with the electric field, by adjusting the length of the probe to the wavelength of the trapped mode. It is critical to choose the position of the coupler precisely where the electrical field is maximum and this is difficult, because the electric field gradients are really high.

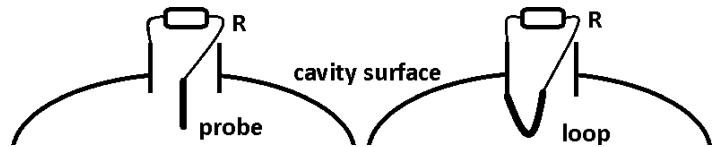

Figure 4: Probe and loop coupling configurations.

Contrarily to the probe HOM coupler, the HOM coupler in the loop configuration interacts with the magnetic field. This configuration allows to reach a greater quantity of field, without wavelength tuning, thanks to the large surface of the antenna exposed to the field.

In general, the probe configuration is useful when the position of the field is well known, while the loop is used when the position of the resonance is not well defined. On the other hand, the loop is less efficient than the probe.

For this reason, one of the main parameters to be identified in the experiment, in order to extract the maximum power from the trapped field, is the position of the HOM coupler, because the coupler should be able to strongly interact with the fields trapped in the resonance.

## 3. Parasitic resonance reduction method

*3.1. Basic idea*

The proposed HOM couplers are aimed at damping unwanted secondary modes [20], and thus differ from couplers currently used in accelerating cavities, which are designed to resonate at given modes. Contrarily to cavities, any other equipment is not designed to be a resonator. The resonances to be damped are only due to spatial discontinuities and have much lower Q-factors, and hence are easier to damp. As a consequence, the proposed HOM couplers are much smaller than the ones used in accelerating cavities.

The proposed HOM couplers can damp resonances over a large frequency band, because it is an all-mode coupler. The idea is not to couple to a single frequency with an equivalent of RLC circuit, but to use an equivalent RL circuit such to cover a wider range of frequencies.
This is the key modification that allows the HOM couplers in loop configuration to be implemented, not only inside a cavity, but also in a generic equipment, that has parasitic resonances.

Examples of equipment where this new method could be used are: the Beam Wire Scanners (BWS) [21] and the Beam Gas Ionization monitors (BGI) [22].

*3.2. Method*

A flow chart of the proposed method is shown in Fig. 5. The different phases of the method will be detailed in the following Subsections.

*3.2.1. Preliminary characterization of the device*

Initially, the equipment need to be characterized, without the HOM coupler. In order to do this, it is necessary to define a model of the device in terms of geometrical and electromagnetic properties of the device under test. After that, it is possible to proceed to estimate the resonance frequencies of the device, the position where the field has its maximum and the value of Q-factor. This should be obtained by both a simulation and an emulation phases. The simulation gives the opportunity to identify the maximum field positions in an easier way than the emulation, as if done by emulation several measurements in different positions should be needed. On the other hand, the emulation allows to obtain more accurate estimations of the Q-factor and of the resonance frequencies than the simulation as the simulation could be affected by a non–perfect modeling of the device properties. The simulation could be carried out by Finite Element Method (FEM) analysis. The emulation should be carried out by the probe method, consisting of measuring the reflection coefficient $S_{11}$, through an antenna placed inside the device.

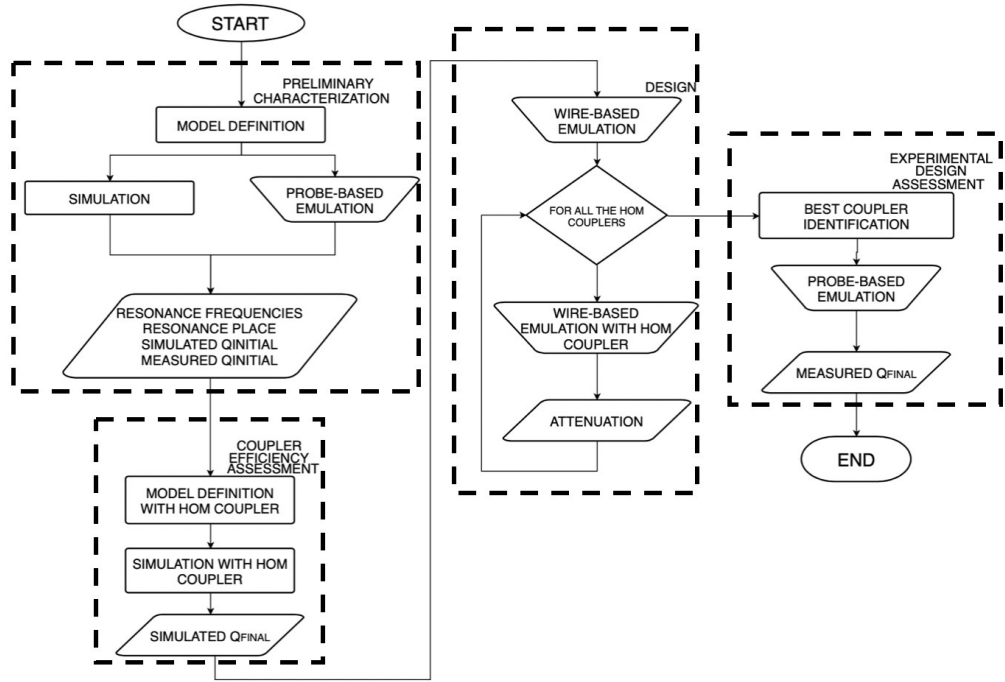

Figure 5: Flow-chart of the methodology.

*3.2.2. Coupler efficiency assessment*

Then, is necessary to verify the capability of extracting energy from the resonance of the considered device, by the use of HOM couplers and thereby reducing the Q-factor.
To do that, a model of the device with the addition of an HOM coupler should be considered. As in the previous case, the model is defined in terms of geometrical and electromagnetic proprieties of the considered device with the addition of the HOM coupler. Then, a FEM simulation should be carried out with the aim of evaluating the reduction of the Q-factor obtained thanks to the insertion of the HOM coupler.

*3.2.3. Design*

Once verified the capability of HOM couplers of effectively reducing the $Q$-factor of the resonances of considered device, the actual design phase can start. A preliminary wire emulation of the empty structure should be carried out, such to determine a baseline of the transmission coefficient, without the HOM coupler. The wire method consists of placing a wire in the middle of the device and measuring the transmission coefficient $S_{12}$. In this phase the wire emulation is preferred to the probe emulation as it is faster. However, it cannot be used to accurately evaluate the Q-factor, as the presence of the wire modifies the boundary conditions and provides a conductive path, helping the EM field to escape the structure. As a consequence, the measured $Q$-factor would be lower than the actual one.
Several HOM couplers with different configurations should then be realized and tested to evaluate the configuration providing the largest attenuation of the resonance peak. In particular, four parameters need to be tuned: the transverse size of the coupler, the length, the number of turns, and the load resistor. Therefore, for each configuration, a wire emulation is performed and the attenuation of the $S_{12}$ peaks, with reference to the baseline.

*3.2.4. Experimental design assessment*

The procedure described in the previous Subsection allows to select the configuration, providing the best result, that will be used further. Therefore, it is necessary to experimentally validate the device with the addition of the HOM coupler in the selected configuration. At this aim, another probe emulation allows to assess the Q-factor reduction due to the introduction of the HOM coupler. For this purpose, a coupler with the selected configuration is placed in each position where a local maximum of the field was found in the preliminary characterization of the device.

## 4. Experimental case study on the Quattro-Tank

The best HOM couplers have to be designed and placed by taking into account the mechanical constrains and the electromagnetic properties of the device. In the following, the proposed method is illustrated for a test equipment with a specific spatial discontinuity, able to cause resonances.

*4.1. Quattro-Tank*

The QuattroTank (QT, Fig. 6) is a Super Proton Synchrotron (SPS) component used as a testbed for crystal collimation [23]. The idea of crystal collimation is to insert a crystal at a given distance away from the beam, usually $3\sigma$ (where $\sigma$ is the transverse beam size). As a consequence, all particles outside $3\sigma$ will hit the crystal and, since the crystal has a circular shape, will follow the crystal structure and be will be deviated away into a collimator. The QT has a geometrically simple structure and thus it is easy to simulate. It has a clear discontinuity of the cross section that will create resonances. It is also made of low resistance material, which will not damp any of the resonances. For these reasons, the QT is an ideal test structure for developing the new method to damp the resonances.
On this device, a full case of study was performed, following the steps described in the previous Section.

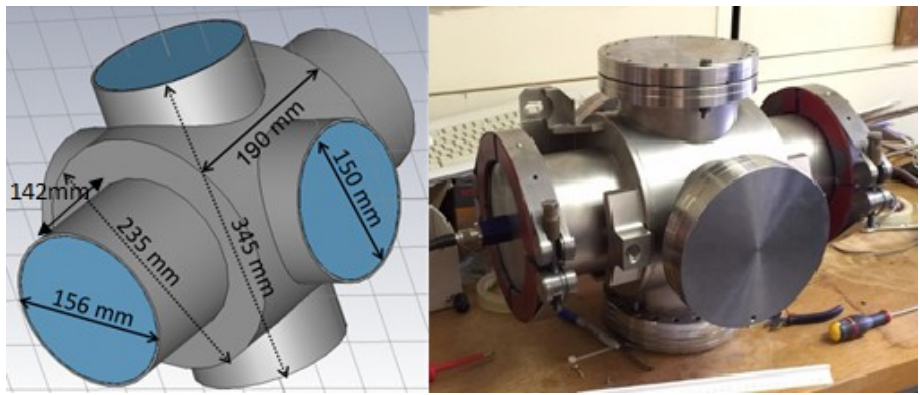

Figure 6: The $3D$ model of the QT (left) and a picture (right).

## 5. Numerical results

Simulations were carried out with two aims: (1) to evaluate the field pattern trapped inside the device, and (2) to assess the $Q$–factor. For these twofold aims, two simulation campaigns were carried out: *Wakefield* and *Eigenmode*. The *Wakefield* simulations are carried out by inserting a beam in the center of the structure and thus evaluating the field pattern inside the cavity while the beam travels through the structure. The Fourier transform of the Wakefield (which is basically an impulse response) is equal to the impedance of the equipment [24]. The *Eigenmode* simulation is an analysis in terms of internal field for each natural resonance mode. Finally, the resonance characteristic ($Q$-factor of the mode) of the structure is obtained.

In all the simulations, the structure is modeled to be composed of lossy material, as it is in the reality, and all the boundaries are grounded, in order to avoid unwanted reflections. The resonances found in the equipment are below the cut-off frequency of the beam pipe, and for this reason they can not escape the equipment.

*5.1. Empty structure simulation*

In the two cases (Wakefield and Eigenmode simulations), two different types of mesh were used. For the Wakefield simulations the largest cell is $2.708\,33$ mm, the smaller is $1.500\,00$ mm for a total of $2\,927\,232$ mesh cells. For the Eigenmode simulations, a tetrahedral mesh type was chosen, the total number of tetrahedrons used was $93\,521$.

The results of the Wakefield simulations are shown in Figs. 7 and 8.

Fig. 7 shows the beam impedance of the structure versus the frequency. It can be observed that the beam excites two resonances in the structure, at $0.920$ GHz (case A) and at $1.320$ GHz (case B). Fig. 8 shows the location of the magnitude of the magnetic field for the two resonances. It is possible to appreciate the regions (in green), showing the highest magnitudes of the field, caused by the resonances. The positions having the highest field magnitudes are those candidates for the placement of the HOM couplers, such that the greatest amount of energy could be absorbed.

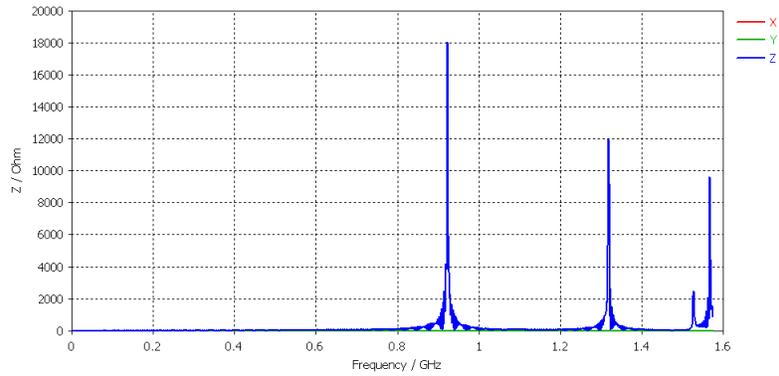

Figure 7: The impedance of the QT, obtained in simulation. The three colors show the impedance in the three plans. The Z plane is the beam plane.

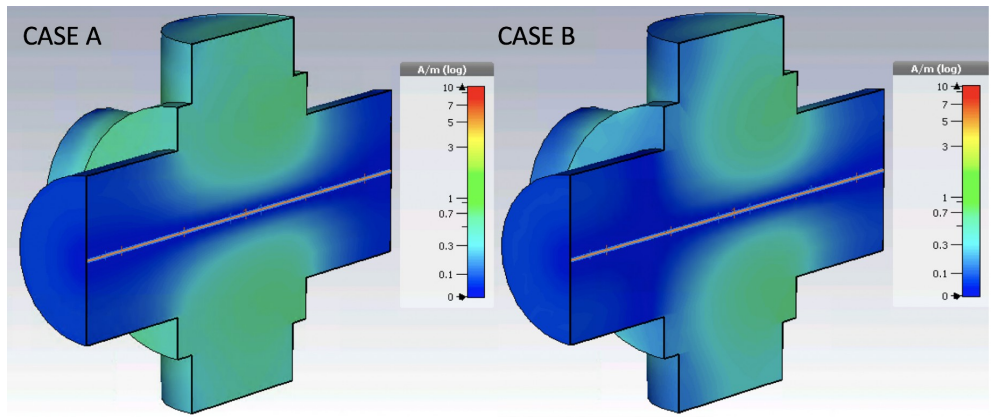

Figure 8: The field magnitude distribution obtained during the EM simulation of the QT for the resonance peak at 0.920 GHz (Case A) and at 1.320 GHz (Case B).

The Eigenmode simulation gave a $Q$-factor value of $1.256 \times 10^4$ for the resonance at 0.920 GHz and $1.310 \times 10^4$ for the resonance at 1.320 GHz.

*5.2. HOM coupler simulation*

In the simulations with the HOM coupler, the mesh grid is an important parameter that needs to be precisely chosen. The reason is that the HOM coupler is made of thin wire, so it is necessary to have a really fine mesh grid. For the Wakefield simulations, the largest cell is 3.798 97 mm, the smaller is 0.600 00 mm for a total of 4 598 289 mesh cells. For the Eigenmode simulations, a tetrahedral mesh type was chosen, the total number of tetrahedrons used was 180363.

In both simulations, the coupler was made of lossy metal, such to emulate the load resistance. In the real case the coupler will be made by a good conducting wire and the caught energy will be dissipated on the load resistor, which is outside the equipment.

In Fig. 9, the structure with the coupler is shown, where it is possible to see the positions chosen in order to damp the first resonance.

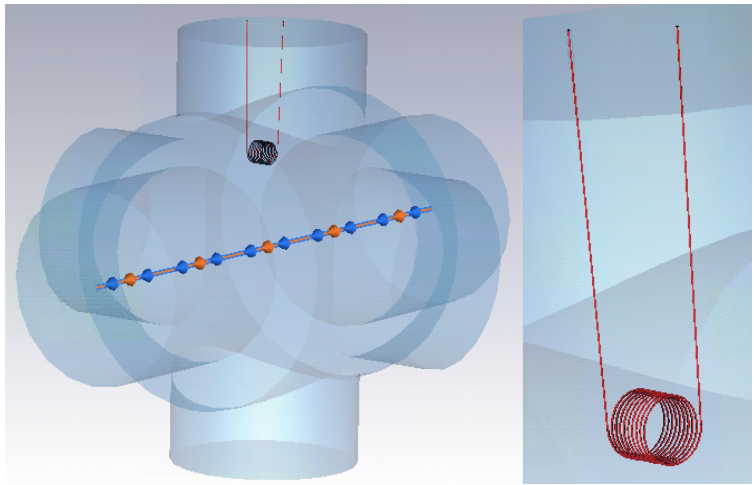

Figure 9: Layout of the simulations with the coupler.

In Fig. 10, the magnitude distribution of the magnetic field is shown, for the two resonances, after the introduction of the coupler. As before, it is possible to appreciate the regions (in green), showing the highest magnitudes of the field, caused by the resonances. Now it is possible to compare the Fig. 8 and 10 and to note the strong reduction of the field.

Thanks to the Eigenmode simulation a *Q*-factor of 979 was obtained for the resonance at 0.920 GHz for the QT with the HOM coupler inside.

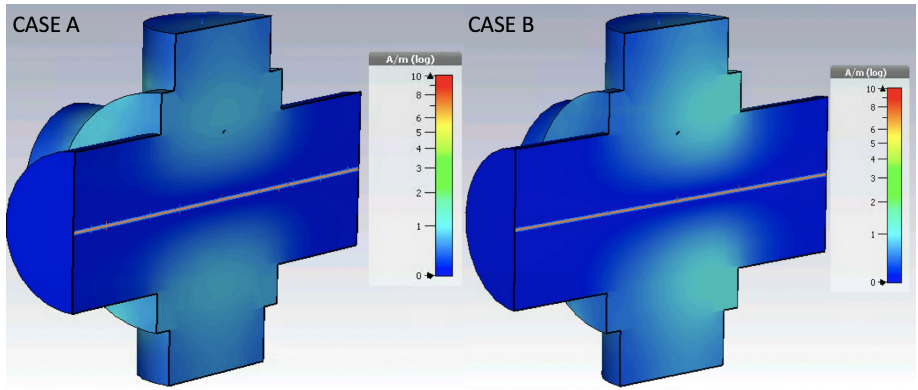

Figure 10: The distribution of the field magnitude obtained by the EM simulation in the case of the QT with the coupler, for the resonance peak at 0.920 GHz (Case A) and at 1.320 GHz (Case B).

Fig. 11 shows the beam impedance versus the frequency of the structure with the HOM coupler inside. Comparing with Fig. 8 it is possible to appreciate the reduction of the impedance peaks at both frequencies. In particular, without the HOM coupler the value of the beam impedance at 0.920 GHz was 18 066 $\Omega$ while with the HOM coupler it is 8722 $\Omega$.

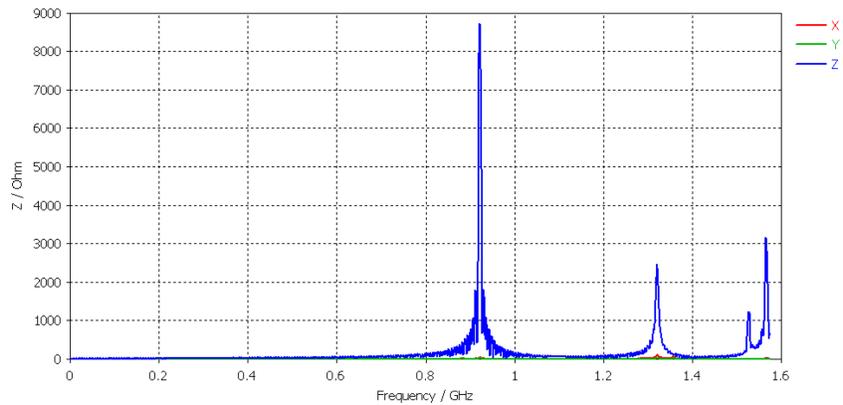

Figure 11: The impedance of the QT with the HOM coupler, obtained in simulation. The three colors show the impedance in the three plans. The Z plane is the beam plane.

## 6. Experimental results

In order to find the best design, several couplers were tested. A total of 27 configurations were selected by varying the diameter (1 cm, 1.4 cm and 1.8 cm), the length (3 cm, 6 cm and 7 cm) and the number of turns (small, medium and large).

The range of values of the above mentioned parameters were chosen such to fit into the equipment without being too close to the beam trajectory or to the equipment walls. The load resistor, used to dissipate the power absorbed by the coupler, was realized by means of a potentiometer, such to tune the resistance value, in order to maximize the transferred power.

The used potentiometer is a Bourns 3339P-1-102, with maximum resistance value of 1 k$\Omega$ and tolerance of 10 % [25]. The wires connecting the coil to the potentiometer were twisted, such to avoid unwanted magnetic coupling.

As already mentioned, two different measurement setups are needed for the evaluation of the experimental analysis. In the next subsections, the two measurement setups will be described in detail and the obtained results will be reported. For both measurement setups a Vector Network Analyzer (VNA) is required, in our case Keysight E5071C was used. The VNA was calibrated in the range of interest 10 MHz to 3 GHz, by means of a 85092C calibration kit, before the use. As results from the VNA data-sheet, the amplitude measurement uncertainty is of 0.05 dB for the transmission measurements and 0.04 dB for the reflection measurements.

### 6.1. Wire Measurement

The wire setup consists of a wire passing through the DUT in expected position of the beam. In this way, the EM field driven by the current in the wire, emulates the behavior of the beam. By means of the VNA, a sinewave signal, whose frequency sweeps up to 1.600 GHz, is then injected into the wire and the transmission coefficient $S_{12}$ is measured for each frequency. A negative peak in the $S_{12}$ graph at a specific frequency corresponds to a resonance trapped into the equipment i.e. an absorption of energy.

The setup is shown in Fig. 12. It is important to underline the necessity of the matching resistors, to avoid the mismatch in the connection between the coaxial cables and the equipment (that is realized by means of Huber-Suhner 66_CB-35-0-1/E SUCOBoxes [26]).

In Fig. 13, the $S_{12}$ coefficient obtained by the wire measurement is shown for the QT without any coupler.

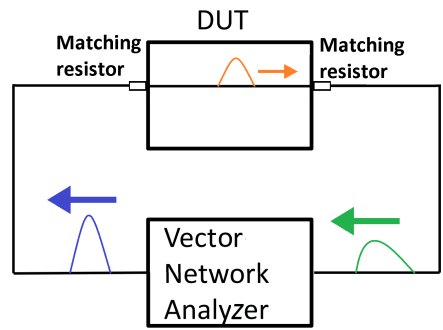

Figure 12: Wire measurement setup used in the experimental approach.

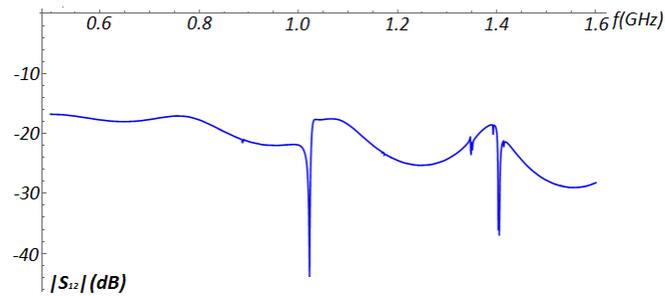

Figure 13: Measurement of $S_{12}$ obtained by wire measurement of the empty QT.

The graph shows clearly 2 main resonances. The first is at $f_1 = 1.018$ GHz, while the second one is at $f_2 = 1.400$ GHz.

These frequencies are slightly different with those obtained by the simulations, shown in Fig. 7, the difference is due to a non perfect matching between the material properties in the simulation and the actual one.

Tests were carried out for all the realized coupler configurations, and the results, obtained in the case of the resonance at the lowest frequency, are reported in Tab. 1.

In the table, the reduction (in dB) of the resonance magnitude peak is also reported. This gives a rough estimate of the suitability of each HOM coupler configuration. The greater attenuation was obtained for the configuration *N. 9*, which is characterized by a diameter of 1.8 cm (the biggest one) a length of 3 cm (the shorter one) and a high number of turns.

Such configuration was selected to be used in the next evaluation steps. Therefore, in a first phase, two couplers with such configuration were placed in the QT, in the positions with maximum magnetic field according to the simulations (see Fig. 8). A new wire measurement test was then carried out. In Fig. 14 the transmission coefficient $S_{12}$, obtained in the second phase, (in blue) is compared with that obtained without the couplers (in red). It can be observed an attenuation of 17.50 dB for the first peak and of 13.92 dB for the second one. Both the peaks were attenuated, although the tested configuration was selected by looking at the attenuation only of the peak at 1.018 GHz.

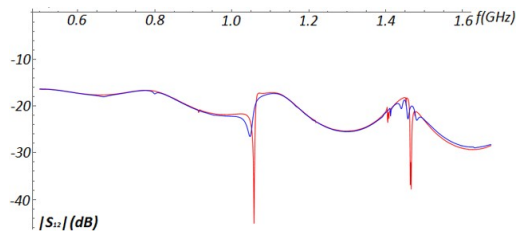

Figure 14: Measurement of $S_{12}$ obtained by wire measurement. In red without the couplers in blue with the couplers.

*6.2. Probe emulation*

The selected configuration, according to the wire measurements in Section 6.1, was tested by means of a probe emulation. In this case, the port 1 of the VNA was connected to an EM probe (shown in Fig. 15) and inserted into the QT (test setup in Fig. 16). The reflection coefficient $S_{11}$ is then measured versus the frequency. The probe was realized at CERN and has a frequency range from 1 MHz to below 10 GHz. The probe radiates an EM field at certain position and senses the field reflected by the DUT. The $S_{11}$ was used in a post-processing step to obtain the Q-factor, as described in [27]. The resulting values were of 2280 without the coupler and 848 with the coupler, for the resonance at 1.400 GHz.

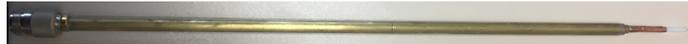

Figure 15: The probe used in the probe method to measure the Q-factor.

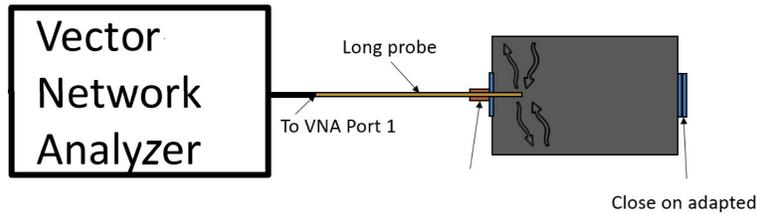

Figure 16: Probe emulation setup.

## 7. Conclusion and future development

In this paper, a new method using HOM couplers for damping resonances inside particle accelerator components has been proposed. HOM couplers in the proposed configuration can be implemented not only inside a cavity but in any generic equipment having parasitic resonances.

The use of the proposed HOM couplers leads to several advantages:

- this method does not require a redesign of the equipment, it does not involve expensive modifications, such as coating, and the price of a coupler is negligible compared to the average equipment used in CERN's laboratories;

- after an EM simulation, it is easy to understand where the coupler should be placed;

- the couplers are not brittle like the ferrite blocks and are also insensitive to temperature;

- the HOM coupler has a broad frequency response, allowing to use the same configuration for different resonance frequencies.

A design procedure for the HOM coupler was also proposed and used in a case study, involving an accelerator component with geometrical discontinuities, and thus causing resonances. Both simulations and experimental results showed the capability of the proposed device to attenuate the resonance peaks and to reduce the Q factor of the resonances. Such results were achieved by means of HOM coupler samples realized in laboratory. Even better results are expected with factory produced samples. In the future, it is planned to apply the proposed method to the Beam Gas Ionization (BGI) [22] monitors that are used in the LHC for measuring the transverse profile of the beam. The BGI monitors are overheating, due to the resonances inside.

## 8. Acknowledgment

The authors thank Elias Metral and Hadron Synchrotron Collective (HSC) section as a whole for useful suggestions, Fritz Casper for highlighting measurement theory, and Kay Papke for the help in the simulation and general theory for couplers. Finally, the authors gratefully acknowledge the CERN Technical Student Programme for supporting the grant of Antonio Gilardi.

| #  | Diameter (*cm*) | Length (*cm*) | Intensity of turns | Attenuation (dB) |
|----|-----------------|---------------|--------------------|------------------|
| 1  | 1               | 3             | LOW                | 0,88             |
| 2  | 1               | 3             | MEDIUM             | 3,13             |
| 3  | 1               | 3             | HIGH               | 1,02             |
| 4  | 1,4             | 3             | LOW                | 1,69             |
| 5  | 1,4             | 3             | MEDIUM             | 1,68             |
| 6  | 1,4             | 3             | HIGH               | 3,95             |
| 7  | 1,8             | 3             | LOW                | 7,99             |
| 8  | 1,8             | 3             | MEDIUM             | 7,43             |
| 9  | 1,8             | 3             | HIGH               | 18,55            |
| 10 | 1               | 6             | LOW                | -0,36            |
| 11 | 1               | 6             | MEDIUM             | 2,63             |
| 12 | 1               | 6             | HIGH               | 3,76             |
| 13 | 1,4             | 6             | LOW                | 1,40             |
| 14 | 1,4             | 6             | MEDIUM             | 6,53             |
| 15 | 1,4             | 6             | HIGH               | 5,83             |
| 16 | 1,8             | 6             | LOW                | 7,78             |
| 17 | 1,8             | 6             | MEDIUM             | 4,77             |
| 18 | 1,8             | 6             | HIGH               | 3,44             |
| 19 | 1               | 7             | LOW                | 8,98             |
| 20 | 1               | 7             | MEDIUM             | 11,80            |
| 21 | 1               | 7             | HIGH               | 7,37             |
| 22 | 1,4             | 7             | LOW                | 13               |
| 23 | 1,4             | 7             | MEDIUM             | 11,03            |
| 24 | 1,4             | 7             | HIGH               | 8,42             |
| 25 | 1,8             | 7             | LOW                | 8,28             |
| 26 | 1,8             | 7             | MEDIUM             | 9,78             |
| 27 | 1,8             | 7             | HIGH               | 1,39             |

Table 1: Measurement results obtained by the wire for the identifications of the best suitable coupler, placed at the position of the lower resonance.